# Application of Selective Algorithm for Effective Resource Provisioning In Cloud Computing Environment

Mayanka Katyal[1] and Atul Mishra[2]

[1]Deptt. of Computer Engineering, YMCA University of Science and Technology, Faridabad, Haryana, India
[2]Deptt. of Computer Engineering, YMCA University of Science and Technology, Faridabad, Haryana, India

## ABSTRACT

*Modern day continued demand for resource hungry services and applications in IT sector has led to development of Cloud computing. Cloud computing environment involves high cost infrastructure on one hand and need high scale computational resources on the other hand. These resources need to be provisioned (allocation and scheduling) to the end users in most efficient manner so that the tremendous capabilities of cloud are utilized effectively and efficiently. In this paper we discuss a selective algorithm for allocation of cloud resources to end-users on-demand basis. This algorithm is based on min-min and max-min algorithms. These are two conventional task scheduling algorithm. The selective algorithm uses certain heuristics to select between the two algorithms so that overall makespan of tasks on the machines is minimized. The tasks are scheduled on machines in either space shared or time shared manner. We evaluate our provisioning heuristics using a cloud simulator, called CloudSim. We also compared our approach to the statistics obtained when provisioning of resources was done in First-Cum-First-Serve(FCFS) manner. The experimental results show that overall makespan of tasks on given set of VMs minimizes significantly in different scenarios.*

## 1. Introduction

Today due to consistently changing requirements of users and increase in demand of services and applications requiring huge processing power and heavy resources has led to the development of cloud computing technology wherein all high cost infrastructure and computational resources are installed in single datacenter. Cloud Computing uses SOA (Service Oriented Architecture) to provide IaaS (Infrastructure as a Service) [1],[2],[3], SaaS (Software as a Service) [1],[2],[3], PaaS (Platform as a Service) [1],[3], DaaS (Data Storage as a Service) [1][2], CaaS (Communication as a Service) [2],[3], HaaS (Hardware as a Service) [1],[2] to cloud users. The end users can use these resources over a network on-demand basis in pay-as-you-say manner.

     1

Resource Provisioning is an aggregation of both Resource allocation and Task Scheduling. "Resources available on the cloud must be provisioned in such a manner that their tremendous capabilities are efficiently utilized and effectively available to end users on-time without much delay in completion of tasks given by the cloud users. The Provisioning must qualify service level requirements (Requirements specified in Service Level Agreement, SLA) based on availability, capability, performance, and costs of resources."

Resource provisioning must take into account the consideration of all kind of cloud users. It must fulfill the requirements of all its users (i.e. the provider and end user of cloud services). Therefore, while designing a resource provisioning technique the developer must emphasize on the needs of Cloud Users on one hand and Cloud Provider on the other hand. The pre-requisite of cloud users is minimum response time. They expect their jobs to be completed in fastest possible manner with high availability of resources. On the other hand the Cloud Provider invests such huge capital with the aim to maximize the use of all the installed resources in efficient and effective manner. Both Cloud users and Cloud providers expect maximum throughput and improved performance for the money they invest. Cloud users pay to use the services of cloud and cloud providers invest huge amount in installation and maintenance of Cloud.

In this paper, we discuss allocation of resources to tasks using min-min or max-min algorithm and then scheduling them on either space shared or time shared basis. The paper is organized as follows. In section 2, the related works are discussed. In section 3, our allocation policy is introduced. In section 4, the experimental results are presented and discussed. We conclude this study in section 5.

## 2. Related Work

Resource provisioning in cloud computing environment is done with the main aim of achieving load balancing. Based on various factors like spatial distribution of cloud nodes, algorithm complexity, storage/replication, point of failure etc. different techniques have evolved to provision the resources in balanced manner. The provisioning is done taking into account whether the environment is static or dynamic.

Many related works have been done to achieve efficient resource allocation scheme. Statistical based Resource allocation (SLB) introduced by Zhenzhong Zhang in paper [4] is based on prior learning and performance statistics, SLB involves analysis of huge on-line historical data for forecasting resource demands. In paper [5], an algorithm given for load balancing is an inspiration from the honeybee. It is biologically inspired technique that uses behavior of honeybees foraging and harvesting for food. It does not take into consideration the waiting time of the tasks and overall turnaround time. Similar techniques mentioned in paper [6], [7] are used for load balancing but none reduce the overall waiting time of tasks.

## 3. Resource Allocation Using Selective Algorithm

To achieve our goal of minimizing the overall makespan of tasks on machines and provide better quality of service we design an algorithm that assigns tasks to best machines in such a way that it provides satisfactory performance to both, cloud users and providers. The algorithm is designed to make a choice between Min-Min and Max-Min scheduling algorithm based on certain criteria.

Min-Min algorithm follows the following procedure:

Phase1: It first computes the completion time of cloudlets on each VM and then for each cloudlet it chooses the VM which processes the cloudlet in minimum possible time.

Phase2: Then among all the cloudlets in MT the cloudlet with minimum overall completion time is chosen and is allocated to VM on which minimum execution delay is expected. The cloudlet is removed from the list of MT and the process continues until MT list is empty.

Max-Min algorithm works similarly like Min-Min in phase 1. Unlike Min-Min, Max-Min allocates the cloudlet with maximum expected completion time to VM (i.e. longer task is first allocated a VM). Max-Min algorithm allocates considerably long task (cloudlet) (long in comparison to other given set of tasks) to one VM and smaller tasks to another VM based on their completion time on each VM. Min-Min on the other hand is useful when all the tasks are almost of same size. When all the tasks are of similar sizes then min-min allocates the cloudlet to a VM on which it executes in minimum possible time. This not only ensures that overall makespan of tasks on VM is reduced but also provide minimum delays in processing of tasks.

To better understand the working of resource allocation and task scheduling algorithms with the aim to achieve load balancing in cloud computing environment a simulated setup was prepared using CloudSim tool.

In our simulation tool CloudSim [8] tasks are modeled as Cloudlets and machines are modeled as VMs and cloud itself is modeled as a Datacenter. VMs are the virtual machines on which cloudlets are executed. These VMs are mapped onto hosts (Physical Machines) based on hardware requirements (processing cores, memory and storage). Processing capabilities of host are measured using MIPS (Million Instruction per Second). Filesize is used as the basic unit to measure the size of the task (cloudlet).

Table 1, gives the sample of execution time of Cloudlets on different VMs in which Max-Min outperforms Min-Min. The tasks are allocated resources using Max-Min and are scheduled in space shared manner. In space-shared scheduling, a given cloudlet is executed to completion before preemption of VM from cloudlet. Whereas in time-shared scheduling more them one cloudlet may be executed on one VM in different time slots by consistent preemption of VMs from cloudlets.

Table1. A sample where Max-Min allocation policy outperforms Min-Min allocation policy

| Tasks / Machines | VM0 | VM1 |
|---|---|---|
| T0 | 1.0 | 0.5 |
| T1 | 2.0 | 1.0 |
| T2 | 3.0 | 1.5 |
| T3 | 9.0 | 4.5 |

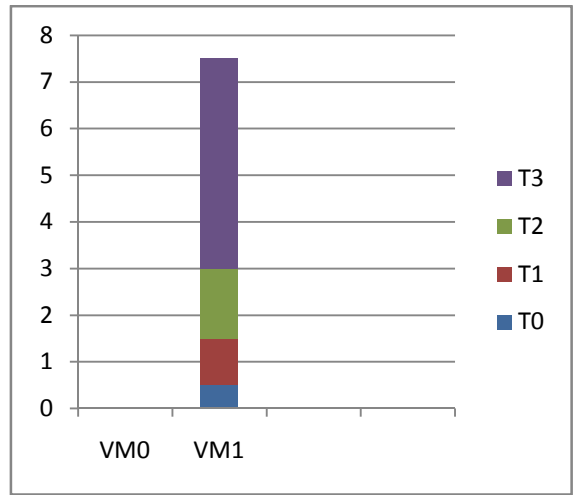

Figure 1. Min-Min (Overall makespan=7.5)

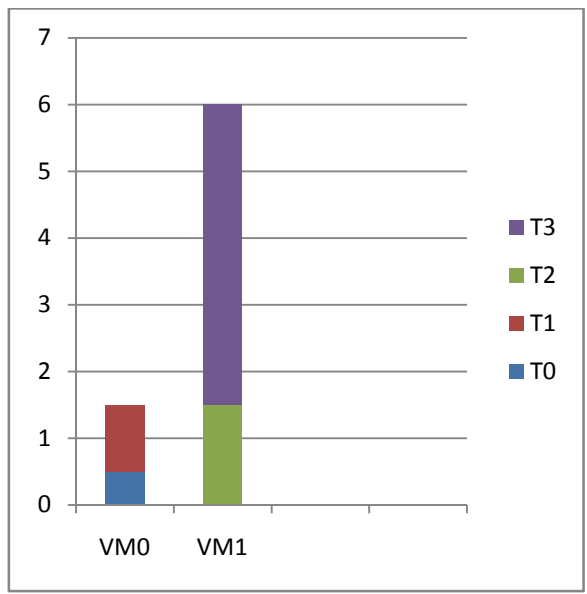

Figure2. Max-Min (overall makespan =6)

Figure 1, shows the overall makespan of four cloudlets on two VMs in space shared manner is 7.5 when Min-Min allocation policy is used. VM0 is underutilized whereas VM1 is over utilized. Since overall completion time of tasks on VM1 is less as compared to VM0, Min-Min allocation technique does not utilize the processing capabilities of VM0. However, in Figure 2. Which uses Max-Min algorithm for resource allocation both the VMs are equally utilized, the load is balanced and overall makespan is reduced to 6. So, in this scenario, Max-Min provides better performance then Min-Min.

Considering the pros and cons of both the allocation policies an algorithm is designed which selects an appropriate algorithm among Max-Min and Min-Min depending upon which gives better performance and minimizes overall Makespan.

```
(1) Input FileSize of all Cloudlets and MIPS of all VMs.
(2) for all cloudlets cᵢ in meta-tasks(MT)
(3)     for all VMs (Virtual Machines) vⱼ
(4)         Exeᵢⱼ = FileSize[i] / MIPS[j]
(5) Sort all the cloudlets in meta-tasks in ascending order of their execution time.
(6) While Meta-tasks list is not empty
(7)     for all cloudlets cᵢ in MT
(8)         for all VMs (Virtual Machines) vⱼ
(9)             Compᵢⱼ = Exeᵢⱼ + Waitⱼ
(10)    for all cloudlets cᵢ in MT
            find the cloudlet with minimum compᵢⱼ and the resource vⱼ that processes it
            in minimum time.
(11)        If more then one resource obtains this minimum
(12)        then
(13)            Select the VM which has been first instantiated on FCFS basis.
(14)    Calculate Standard Deviation (SD). Using relation (4).
(15)    Find location (l) in Meta-tasks (MT) where the difference between two consecutive
        values of compᵢⱼ is greater than standard deviation (SD).
(16)    If 'l' is present in the upper half (l< c/2) or SD < Mean_Compᵢⱼ
(17)    then
(18)        Allocate VM to Cloudlet using Min-Min.
(19)    else
(20)        Allocate VM to Cloudlet using Max-Min.
(21)    Remove allocated cloudlet from MT.
(22) End while.
```

Figure 3. Selective Algorithm

If v denotes the number of VMs and c denotes the number of cloudlets or meta-tasks (tasks which are yet not processed) then Max-Min allocation policy has the time complexity of O ($c^2v$) and Min-Min has time complexity of O ($c^2v$).

In the selective algorithm specified in Figure 3, initially the Filesize of each $i^{th}$ cloudlet and MIPS of each $j^{th}$ VM is given as input and in the first for loop execution time of cloudlets is calculated on each VM is calculated using following relation, lines (1-4):

$$Exe_{ij} = \frac{FileSize[i]}{MIPS[j]} \qquad (1)$$

All cloudlets will be then sorted according to minimum execution time, line (5).

Then in the second for loop, expected completion time of cloudlets on each VM is calculated using the following relation:

$$Comp_{ij} = Exe_{ij} + wait_j \qquad (2)$$

Where $Exe_{ij}$ is the execution time of $i^{th}$ cloudlet on $j^{th}$ VM, $wait_j$ is the time for which $i^{th}$ cloudlet has to wait for $j^{th}$ VM to get ready. This is done in lines (7-9).

In the third for loop, we find the cloudlet with minimum expected completion time from the given set of meta-tasks and the resource that processes it in minimum time, line (10). If more than one resource obtains this minimum then the resource object (i.e. VM) which was first instantiated is chosen in line (11-13).

Now to select between Max-Min and Min-Min, Standard Deviation is used as a heuristics. Standard deviation is calculated in line (14) and the relation used is as follows:

$$Mean\_Comp_{ij} = \frac{\sum_{i=1}^{C}(Comp[ij])}{C} \qquad (3)$$

$$SD = \frac{\sqrt{\sum(Comp[ij] - Mean\_Comp)^2}}{\sqrt{C}} \qquad (4)$$

where $Mean\_Comp_{ij}$ denoted average of $Comp_{ij}$.

Then, the sorted list is searched for a location where the difference between two consecutive values of completion time is greater than standard deviation (SD), line (15). This heuristics is beneficial in finding the location where difference between completion time of two consecutive cloudlets is sufficiently large as compared to other cloudlets. Also the list is divided into two parts, upper half and lower half. Based upon location following methods are invoked:

Case1: If this location is present in the upper half of the list then it implies the list has greater number of sufficiently long task as compared to short tasks (i.e. tasks of similar sizes are greater). In this case Min-Min outperforms Max-Min and hence Min-Min is invoked as in line (18).
Case2: If this location lies in the lower half of the list then it implies that list contains large number of short tasks but fewer long tasks. In this case Max-Min outperforms Min-Min. Hence Max-Min is invoked as in line (20).

Case3: If no such location exist, that implies that all the task are of almost same sizes with no major difference between any two task sizes (i.e. difference < SD). In this case, following conditions are evaluated:

- If (SD < $Mean\_Comp_{ij}$), then the list contains all small size tasks. In this case Min-Min is executed to allocate a VM to cloudlet, line (16).
- If (SD >= $Mean\_Comp_{ij}$), then Max-Min is executed to allocate a VM to cloudlet.

After the allocation of a VM to the cloudlet the cloudlet is removed from the list of MT and the process continues until all cloudlets are allocated to a VM.

## 4. Experimental Results

### 4.1 Performance Metrics:

Makespan of given cloudlets on given set of VMs is used as the performance metrics in this resource provisioning technique in cloud computing environment. Throughput of the heterogeneous system is the measure of the Makespan. It is calculated using the following relation:

$$\text{Makespan} = \max_{c[i] \in MT} (\text{Comp}[i]) \qquad (5)$$

Lesser the value of the Makespan of allocation algorithm better is the resource utilization.

### 4.2 Experimental Testing:

#### 4.2.1 Simulation Environment

To evaluate and compare our allocation strategy with the technique used to allocate resources using FCFS (in the absence of any defined heuristics), a simulation environment has been setup using CloudsSim [8] toolkit. The allocation policy is simulated under following environment:

a) Resource capabilities are defined in terms of MIPS (Million Instructions per Second).
b) Cloudlets (or tasks) are attributed by their FileSize.
c) Resources may be heterogeneous in nature.
d) The environment is static i.e. list of MT and resources and values of their respective attributes is to be given before simulation.
e) The cloudlets can be scheduled in space-shared or time-shared manner on VMs
f) VMs are allocated to host in space-shared manner or time-shared manner.

#### 4.2.2 Experimental Data and Results

The Input Data used in experimental testing is as follows:

Table 2: Input data

| | |
|---|---|
| No. of VMs | 2 |
| No. of datacenter to be created | 1 |
| No. of Hosts | 2 |
| RAM required for VM(in MB) | 512 |
| No. Of Cloudlets(Tasks) | 6 |
| Size of File_0 | 12 |
| Size of File_1 | 16 |
| Size of File_2 | 50 |
| Size of File_3 | 30 |
| Size of File_4 | 20 |
| Size of File_5 | 40 |
| MIPS of VM_0 | 10 |
| MIPS of VM_1 | 20 |

The data given in Table 2 was used to evaluate the performance of our algorithm. The completion time of each task is calculated in the simulated environment. In the given simulation environment, setup using CloudSim, the completion time of each cloudlet is calculated in millisecond and additional time of few milliseconds taken in initialization of each entity of Cloud Computing environment (datacenter, hosts, VMs and cloudlets) is added to overall completion time of cloudlets on their bounded VM (the VM allocated to the cloudlet). The experimental testing of our allocation policy is performed in following scenarios:

Scenario 1: Overall Makespan of six tasks on two VMs in space sharing mode without using selective algorithm), Figure 4(a).

Scenario 2: Overall Makespan of six tasks on two VMs in space sharing mode (using selective algorithm), Figure 4(b).

Scenario 3: Overall Makespan of six tasks on two VMs in time sharing mode is without using selective algorithm), Figure 4(c).

Scenario 4: Overall Makespan of six tasks on two VMs in time sharing mode is (using selective algorithm), Figure 4(d).

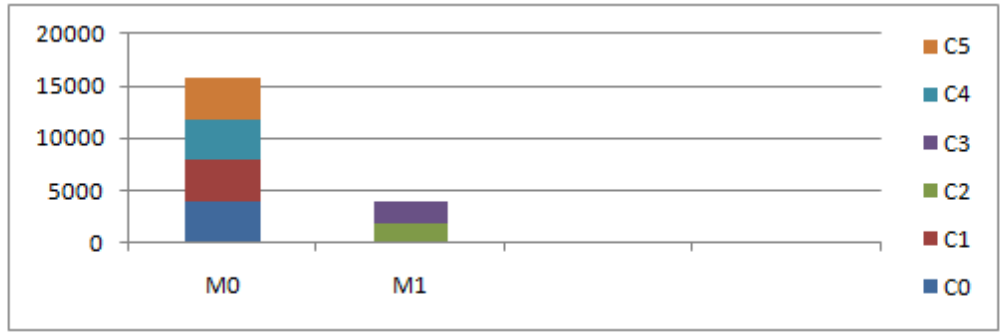

Figure 4(a) Overall Makespan of six tasks on two VMs in space sharing mode(without using selective algorithm).

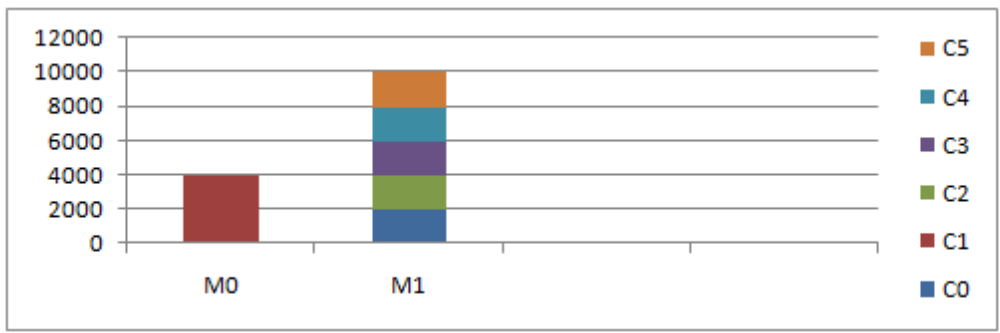

Figure 4(b) Overall Makespan of six tasks on two VMs in space sharing mode (using selective algorithm)

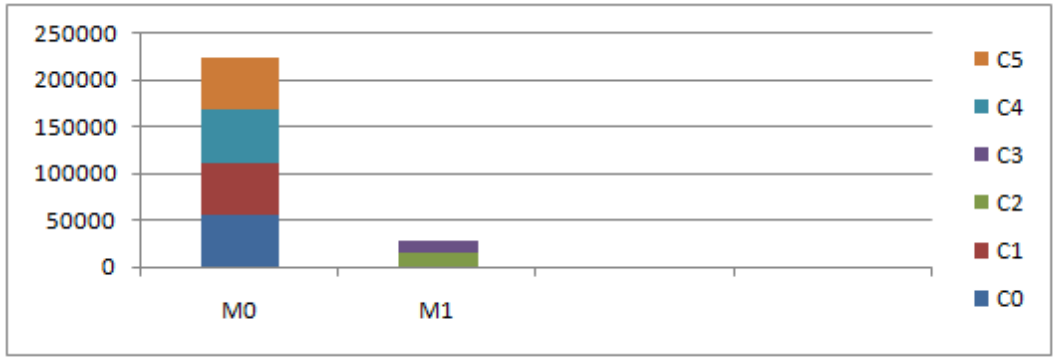

Figure 4(c) Overall Makespan of six tasks on two VMs in time sharing mode is without using selective algorithm

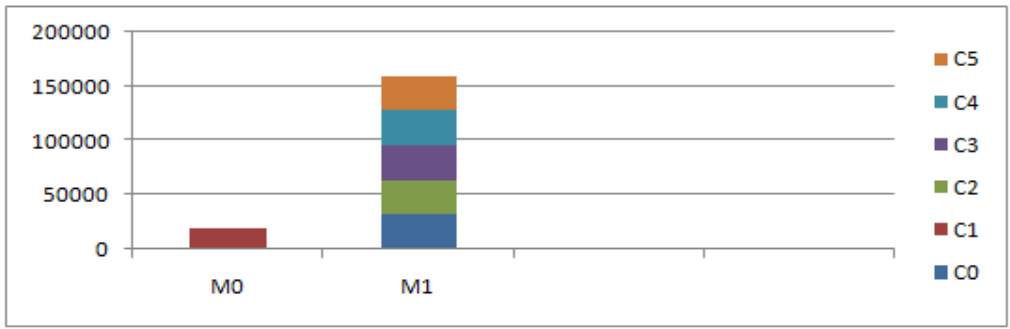

Figure 4(d) Overall Makespan of six tasks on two VMs in time sharing mode is (using selective algorithm)

In Figure 4 (a)-(d), show overall makespan of six cloudlets (C0, C1, C2, C3, C4, C5) on two VMs (M0, M1) in space sharing and time sharing scheduling mode using selective algorithm for resource allocation and comparing it with FCFS based allocation.

Figure 4(b) and 4(d) show that Selective algorithm for resource allocation in cloud computing environment outperforms random allocation in FCFS manner. The overall Makespan of six cloudlets on two VMs using selective allocation policy in space-sharing and time-sharing mode is 10000 and 160000 respectively. Whereas in the absence of appropriate allocation policy the overall Makespan of six cloudlets on two VMs is 16000 and 220000 in space sharing and time-sharing mode respectively, Figure 4(a) and 4(c). Table 3 shows the comparison of four different scenarios simulated in given environment.

Table 3: Makespan of six tasks on two VMs in different modes with and without using Selective Algorithm

| Mode/Algorithm | Without Using Selective Algorithm | Using Selective Algorithm |
|---|---|---|
| Space Sharing | 16000 | 10000 |
| Time sharing | 220000 | 160000 |

Hence, Resource allocation using Selective algorithm outperforms random allocation policy.

## 5. Conclusion

Cloud Computing involves investment of huge capital to install tremendous resources at a single Datacenter. Therefore, to maximize the use of tremendous capabilities offered by Cloud these resources need to be provisioned efficiently. The algorithm given in the paper ensures that all resources are efficiently utilized and jobs given by the user are executed with smaller delays. The provisioning technique given in the paper attempts to improve the throughput by minimizing Makespan. In future, dependencies between tasks can be modeled to further minimize the Makespan and maximize the throughput using Workflows [9] and DAG [10].